# Analemmatic and Horizontal Sundials of the Bronze Age (Northern Black Sea Coast)

**Larisa Vodolazhskaya[1]**

[1] Department of Space Physics, Southern Federal University (SFU), str. Zorge, 5, Rostov-on-Don, 344090, Russian Federation; E-mail: larisavodol@yahoo.com

**Abstract**

The paper presents the results of a study of unique slab with images of Srubna burial of tumuli group Popov Yar-2 (Ukraine) and slab of Srubna burial of tumulus field Tavriya-1 (Russia). A distinctive feature of the images is the orderliness and symmetry of the composition, as well as the location of elliptical cupped depressions. With the help of mathematical and astronomical methods we prove in this paper that the slabs with the images are the ancient sundials. At the Popov Yar-2 slab located two sundials, which worked at the same time: the analemmatic sundial and the horizontal sundial with two gnomons and the linear scale. At the Tavriya-1 slab located analemmatic sundial. On the basis of the reconstruction of the linear parameters of the gnomon of both Popov Yar-2 slab sundials and given the scale value of horizontal sundial, in the article that the potential indirect impact protoscientific knowledge of ancient Egypt to the Srubna population in the Northern Black Sea coast.

**Keywords:** Srubna burial, slab, wells, cupped depressions, analemmatic sundial, horizontal sundial, gnomon

**Introduction**

In the course of archaeological excavations in 2011 in the north-western part of the Donetsk region (Ukraine) in the burial 7 of tumulus 3 of tumuli group Popov Yar-2 was found an unique stone floor slab of the burial on which the images were marked (Fig. 1). Burial and slab are Srubna culture and date XIII - XII centuries BC [1].

On the surface of the slab on both sides were marked the composition of lines and cupped depressions arranged in an oval. Wells, or as they are called, cupped signs are present in the rock art in all continents with the Paleolithic period. However, so far there is no consensus about its interpretation. Sometimes small cupped marks find on the stones of cysts, passage graves, dolmens and ancient cemeteries. For example, in Azerbaijan there is a tradition where the local residents in the spring holiday Nowruz put into such recesses wheat sprouts or sweets as a gift to the spirits of the dead [2].

A distinctive feature of the compositions of the cupped depressions on both sides of the Popov Yar-2 slab is orderliness and symmetry in the placement of cupped depressions and reduced straight grooves. This fact allowed us to assume that the composition of the cupped depressions and the grooves were not only related to solar symbolism [3], but also served to measure time by the sun directly, being marks for sundial. In European countries for a long time, there is an ancient tradition to place on the graves of sundial all the more so [4]. It is known that the image of a sundial was found in the tomb of Seti I (about 1300 BC) in Egypt [5].

**Analyzing image of the groove with transverse lines on the side**

On the two sides of the Popov Yar-2 slab were inflicted reduced straight lines and knocked cupped depressions - wells. For the analysis of the images on the surface of the slab were used high-quality pictures of the report documentation by archaeological excavations (Fig. 1).

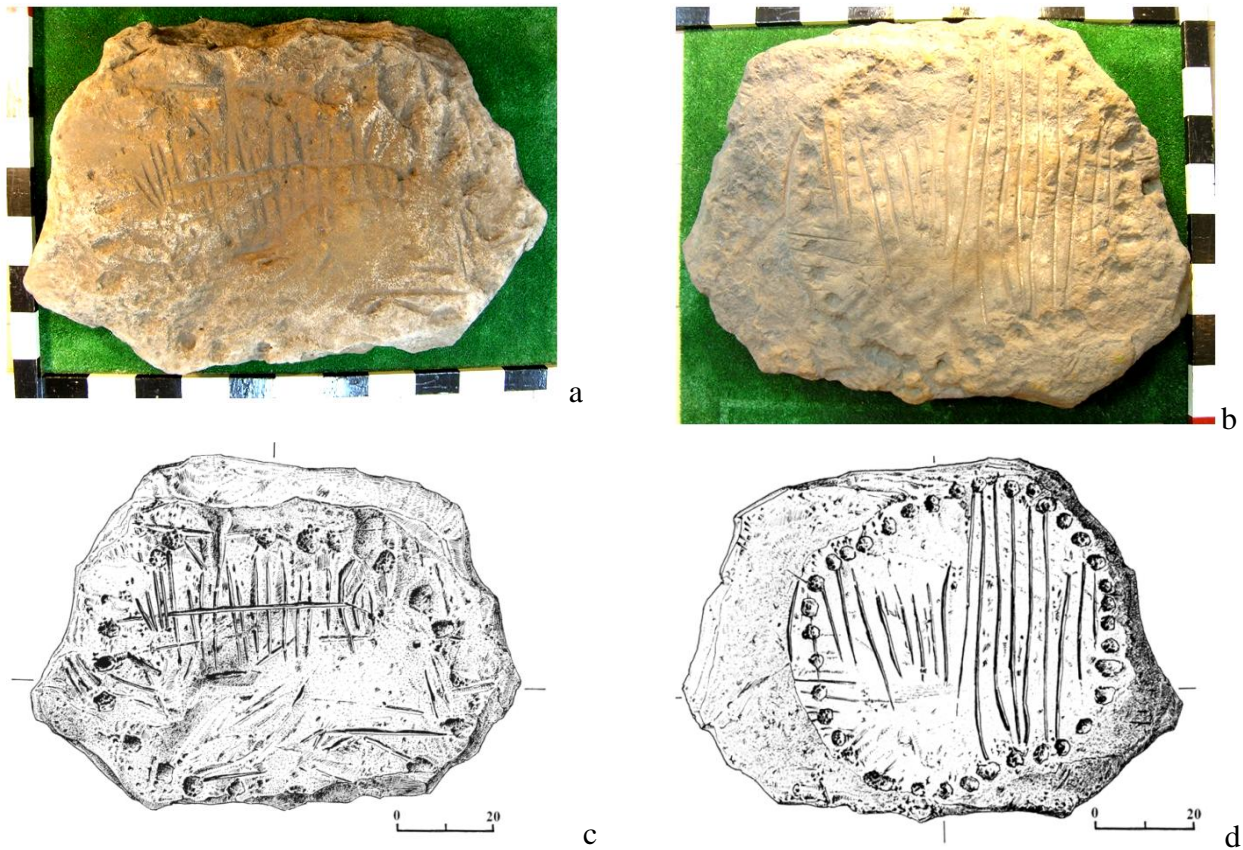

**Figure 1**. Popov Yar-2, tumulus 3, burial 7, slab: **a** - photo of side A [6], **b** - photo of side B [7], **c** - drawing of side A [8], **d** - drawing of side B [8].

The characteristic distortions images were adjusted using the graphical editor Adobe Photoshop CS5 (Fig. 2). The correction carried out using additional photos taken with perpendicular to each other photometers lying horizontally at the upper surface of the plate. As a result, images on photos represent the projection on the horizontal plane that is perpendicular to the line of sight.

On the side A is clearly visible composition from long groove and transverse lines (Fig. 2). We have suggested that the composition was a linear scale of horizontal sundial. Linear scale of horizontal sundial, similar to that found on the Popov Yar-2 slab, can be obtained by transverse lines mark the point of intersection of the gnomon shadow of the long groove at fixed intervals.

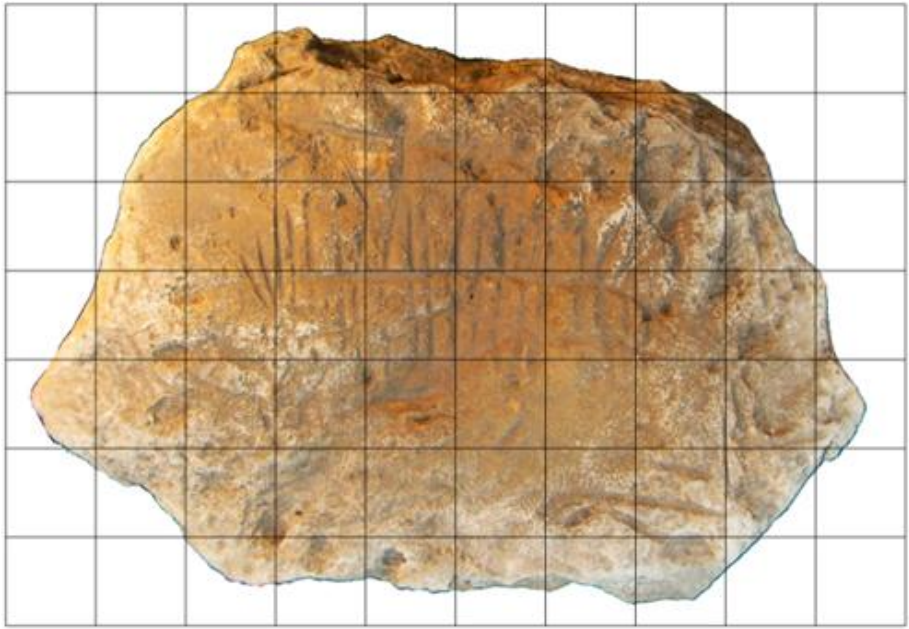

**Figure 2**. Popov Yar-2, tumulus 3, burial 7, slab, the corrected photo of side A.

Fixed intervals can be measured using a water clock. A water clock in the form of vascular with markings inside existed in ancient Egypt since XV - XIV centuries BC exactly [9]. Srubna vessel with the marks on the inside, which could be used as a water clock is filled with the type, was found in the ruined tumulus near the Staropetrovsky village (Yenakiyevo town, Donetsk region, Ukraine). [10, 11, 12] (Fig. 3).

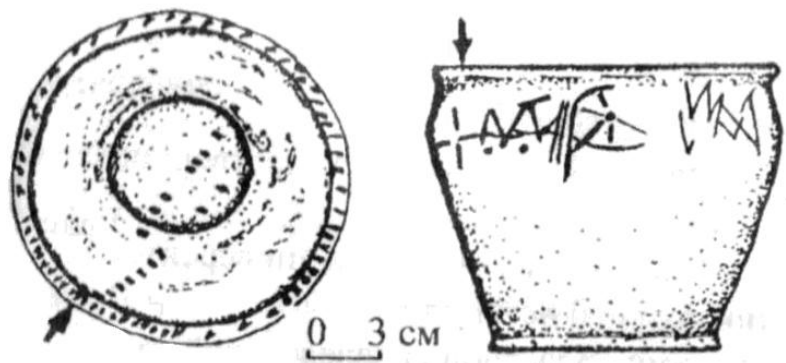

**Figure 3**. Staropetrovsky village, the ruined tumulus, the vessel with the marks on the inside [13].

The calculation of the hour angles for the gnomon of horizontal sundial made under the formula [14]:

$$H' = arctg(\sin\varphi \cdot tgH), \qquad (1)$$
$$H = 15^0 \cdot (t - 12)$$

where H - the hour angle of the Sun (for a half-day $H=0^0$), t - time, $H'$ - an angle between the noon line and the hour line on sundial, $\varphi$ - geographic latitude.

The results of our calculations using the formula 1 for the latitude Lat = 48 ° 26 'N in the time range of from 6 to 18 hours are shown in Tables 1 and 2.

**Table 1**. Hour lines of horizontal sundial (before noon time). *H* - the hour angle of the Sun, *t* - time, *H′* - an angle between the noon line and the hour line on sundial.

|  | t, (hour) | | | | | | | | | | | | |
|---|---|---|---|---|---|---|---|---|---|---|---|---|---|
|  | 6.0 | 6.5 | 7.0 | 7.5 | 8.0 | 8.5 | 9.0 | 9.5 | 10.0 | 10.5 | 11.0 | 11.5 | 12.0 |
| H, (°) | -90.0 | -82.5 | -75.0 | -67.5 | -60.0 | -52.5 | -45.0 | -37.5 | -30.0 | -22.5 | -15.0 | -7.5 | 0.0 |
| H′, (°) | -90.0 | -80.0 | -70.3 | -61.0 | -52.3 | -44.3 | -36.8 | -29.9 | -23.4 | -17.2 | -11.3 | -5.6 | 0.0 |

**Table 2**. Hour lines of horizontal sundial (after noon time). *H* - the hour angle of the Sun, *t* - time, *H′* - an angle between the noon line and the hour line on sundial.

|  | t, (hour) | | | | | | | | | | | | |
|---|---|---|---|---|---|---|---|---|---|---|---|---|---|
|  | 12.0 | 12.5 | 13.0 | 13.5 | 14.0 | 14.5 | 15.0 | 15.5 | 16.0 | 16.5 | 17.0 | 17.5 | 18.0 |
| H, (°) | 0.0 | 7.5 | 15.0 | 22.5 | 30.0 | 37.5 | 45.0 | 52.5 | 60.0 | 67.5 | 75.0 | 82.5 | 90.0 |
| H′, (°) | 0.0 | 5.6 | 11.3 | 17.2 | 23.4 | 29.9 | 36.8 | 44.3 | 52.3 | 61.0 | 70.3 | 80.0 | 90.0 |

The marks on the groove - the scale of the Popov Yar-2 slab are located fairly evenly, which indicates that the simultaneous use of two similar gnomons, fixed at an equal distance from the scale. The traces in the form of a cross located reduced straight lines from the possible fixing the gnomon viewed on the slab at a distance of about 15 cm from the scale (Fig. 4). Using the slab as a sundial determines its orientation in space. Side, where the gnomon, was located in the south. The scale should be in the north of it and be oriented along the axis of the east - west, respectively.

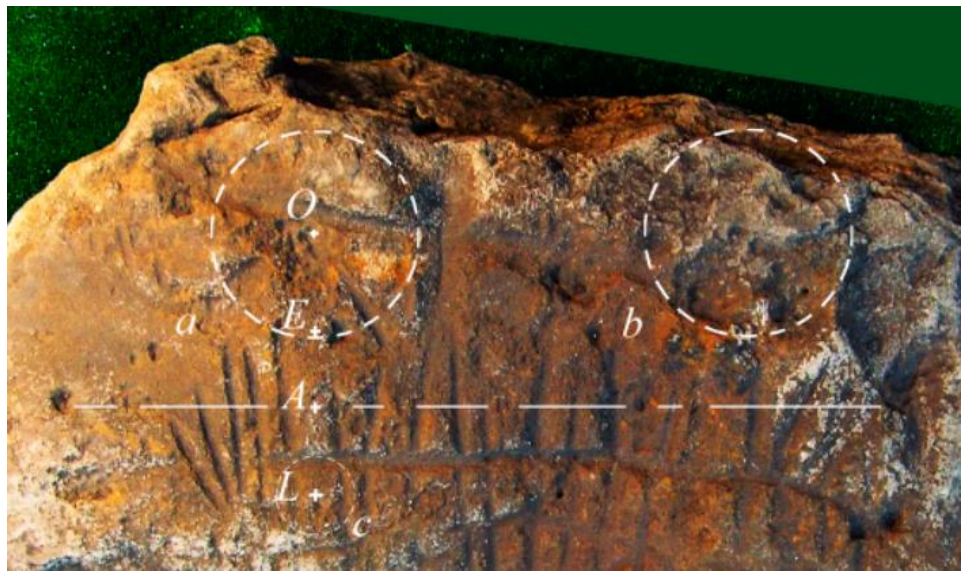

**Figure 4**. Popov Yar-2, tumulus 3, burial 7, slab, side A : a - mounting location the gnomon I, b - mounting location the gnomon II, p - hole corresponding to the 12:00 mark gnomon I, O - the beginning of the gnomon base I, E - end of the base gnomon I, A - the projection on the slab surface upper corner of the gnomon I (projection of the plumb), L – the gnomon I shadow boundary at noon on the summer solstice.

On corrected photo of side A we have marked hour lines of horizontal sundial with the centers in the alleged places of fixing of the gnomons (Fig. 5). In this case, nearly all transverse grooves - risks on the groove - scale quite accurately coincide with the points of intersection of the hour lines whit it (Fig. 6 a). This fact is evidence in favor that a composition of a long groove and transverse grooves - risks was really a linear scale horizontal sundial.

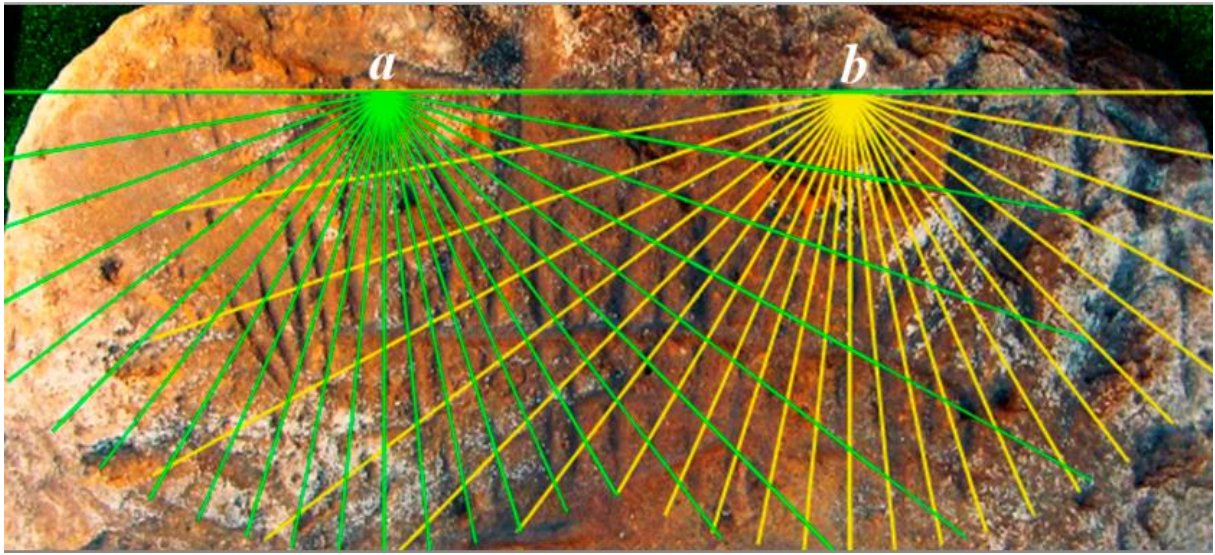

**Figure 5**. Popov Yar-2, tumulus 3, burial 7, slab, side A: **a** – the hour lines of gnomon I, **b** – the hour lines of gnomon II.

The mark of the 12 o'clock of the gnomon I marked no groove, but point the hole near the recess groove. Rather, it was a point of reference when marking scale of the horizontal sundial.

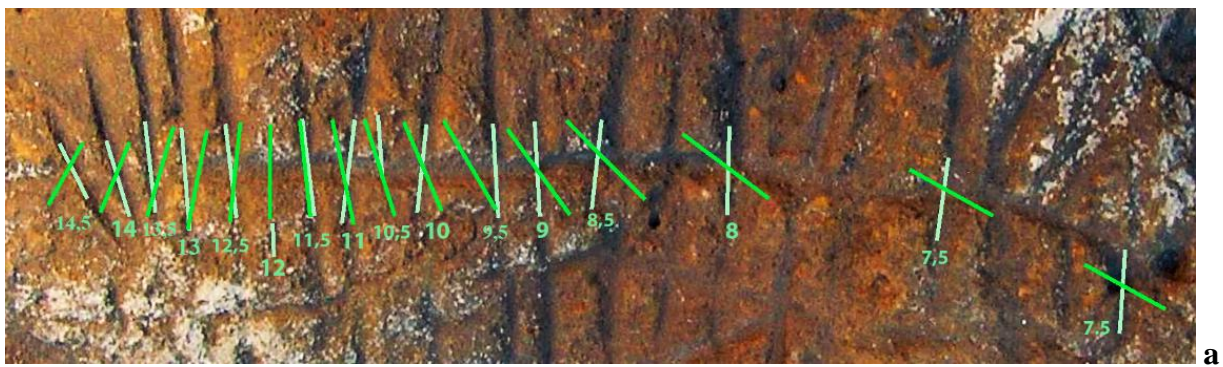

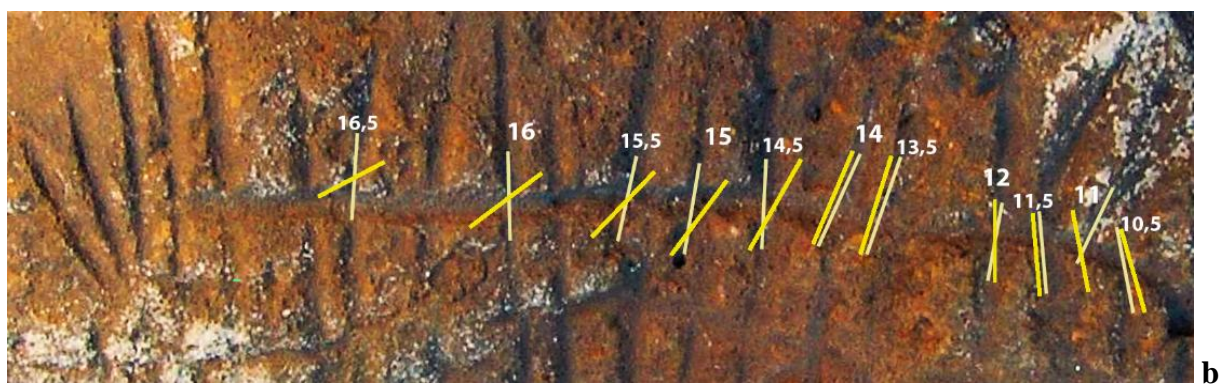

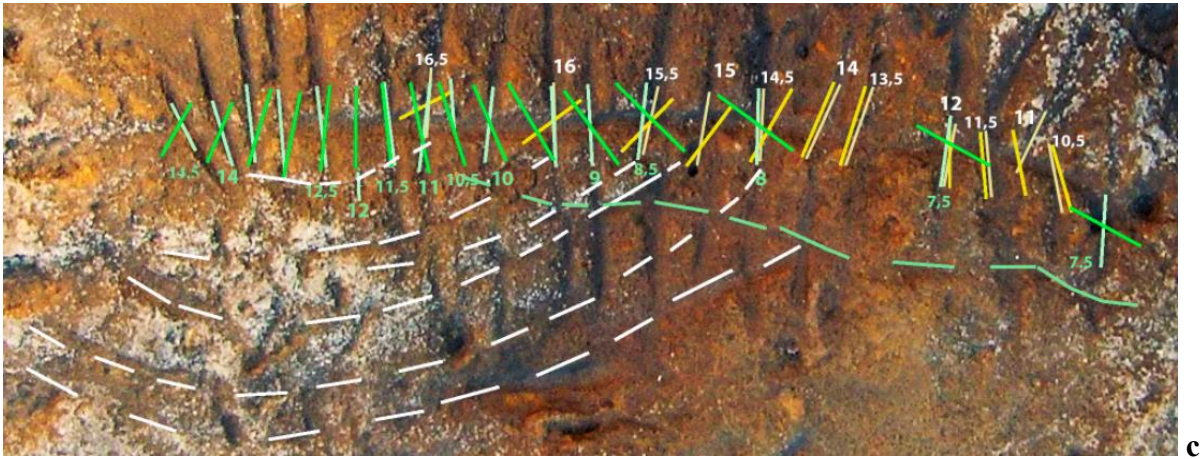

**Figure 6**. Popov Yar-2, tumulus 3, burial 7, slab, side A: **a** - the intersection of the hour lines of gnomon I (green lines) with a groove - scale, **b** - the intersection of the hour lines of gnomon II (yellow lines) with a groove – scale, **c** - reduced straight dotted lines in the north of the groove. Grooves - risks for the gnomon I marked in light green for the gnomon II marked in a light yellow color. Numerals indicate clocks, corresponding to the hour lines.

Grooves - risks indicate the time range of 7.5 to 14.5 hours on a scale for the gnomon I and the range of 10.5 to 16.5 hours for the gnomon II (Fig. 6 b). Thus, to determine the time in the morning used gnomon I, and for the evening time - gnomon II. Using two gnomons allowed halving the length of the linear scale, and gnomons overlapped in time close to noon, allowing the day to monitor the accuracy of its installation.

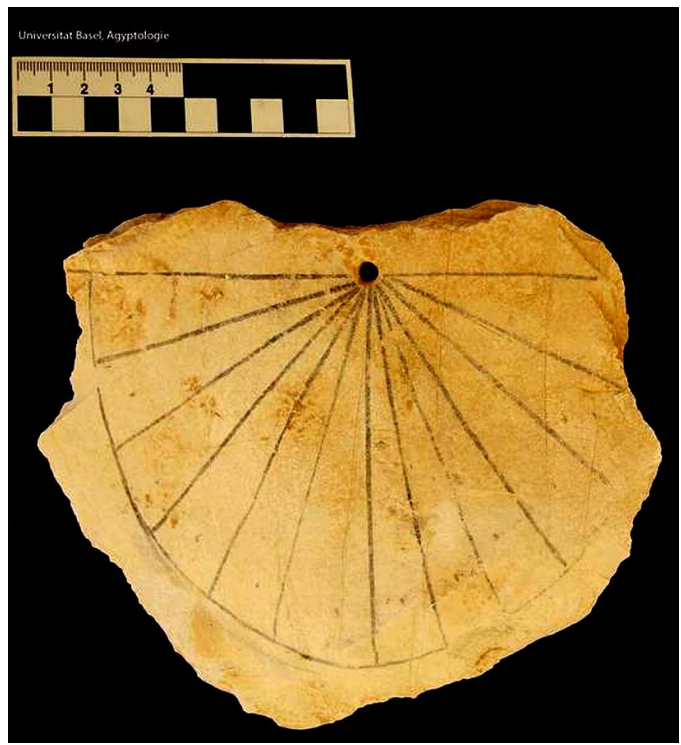

**Figure 7**. Sundial from the Valley of the Kings (Egypt) [1].

---

[1] http://aegyptologie.unibas.ch/forschung/projekte/university-of-basel-kings-valley-project/report-2013/

A characteristic feature of the Popov Yar-2 horizontal sundial is the half-hour marks on the scale. It is believed that division of the clock first appeared in Egypt. Already in 2100 BC. e. Egyptian priests divided the day into 24 hours. The Babylonians also divided the day and night for 12 hours. According to Herodotus: "... with the hemisphere, with the gnomon and the twelve parts of the day the Greeks became acquainted from the Babylonians" (II, 109). About a half-hour division of time in antiquity was not known until recently. However, in early 2013 during archaeological excavations in the Valley of the Kings (Egypt) the University of Basel (Switzerland) expedition have been found sundial dating from the XIII century BC, with point marks, dividing each hour into two parts (Fig. 7).

Exposed to such risks on the scale of the horizontal sundial on Popov Yar-2 slab, corresponding to 12.5 and 13 clocks for the gnomon II. It is known that in Ancient Greece the working day began at dawn and lasted until noon, which marked the end of the working time (Anth. Pa1., X, 43). Dinner time in ancient Rome (Mart., IV, 8), also accounted for the period nearby noon [15]. Lack of the midday risks on the Popov Yar-2 slab sundial of indicates a meal break, possibly also.

**Dashed lines**

Subtle reduced straight dashed lines are viewed in the north from scale on surface of the slab (Fig. 6c). They could be auxiliary markers to denote of hours. The dashed line marked green color in the figure, is an auxiliary mark to emphasize the range of the gnomon I hour lines: from 7.5 to 10.5 hours. Only this gnomon could measure time in this range. White color indicates the dashed line, which are ancillary to time lines of the gnomon II. The dashed lines marked white color in the figure. They are ancillary marks to the gnomon II hour lines. Each such line begins at the mark-risk in the range of 14 to 16.5 hours in which the gnomon II could measure the time only (except for the marks of 14 and 14.5 hours - the extreme marks to gnomon I). Auxiliary lines were equivalent numerical signatures for grooves - risks.

Range from 7.5 to 16.5 hours corresponds to the daylight in the shortest days of the year - the winter solstice, approximately. Daylight is composed of daytime when the sun is above the horizon, and the time of civil twilight, when the angle of the sun dipping below the horizon is less than $6^0$. The sun is above the horizon for about 8 hours, and civil twilight lasts about 40 minutes in the morning and evening[2], for the geographical coordinates of Popov Yar-2 during the winter solstice.

Detailed marking of the scale of horizontal sundial in the range of 7.5 to 16.5 hours associated with the tradition of the working day duration limiting to minimum daylight in a year, maybe.

**Reconstruction of the linear parameters of the horizontal sundial gnomons.**

**Marks of equinoxes and solstices**

Calculations of the gnomons sizes were made based on the need to work horizontal sundial throughout the year, including the summer solstice, when the shadow at noon is the shortest. As a reference point has been chosen different from all other risks as well, corresponding to 12 hours for the gnomon I. The measured distance from the proposed site of gnomon I attachment to this point $l_{ss} \approx 18$ cm (Figure 3 c). During the summer solstice, the sun declination $\delta$ equals obliquity of the ecliptic (the angle of inclination of the ecliptic to the celestial equator) $\varepsilon$, which is calculated using Formula 2 and Formula 3 [16]:

$$\varepsilon = 23.43929111^0 - 46.8150'' \cdot T - 0.00059'' \cdot T^2 + 0.001813 \cdot T^3 \qquad (2)$$

$$T \approx \frac{(y - 2000)}{100}, \qquad (3)$$

where T - the number of Julian centuries separating the era from noon January 1, 2000, y – the year. During the winter solstice the sun's declination $\delta = -\varepsilon$, and at the equinox declination $\delta = 0$. Calculated by us to 1200 BC the formulas 2 and 3 obliquity of the ecliptic $\varepsilon = 23,84^0$.

---

[2] Calculations of day length and duration of civil twilight were performed using the astronomical program RedShift 7.

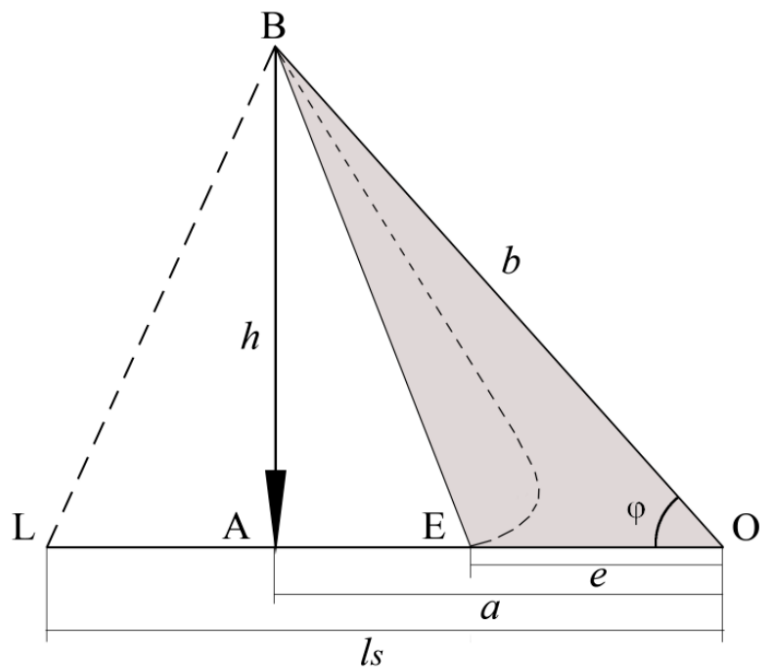

**Figure 8**. Triangle *OBE* – the gnomon (the possible border gnomon more complex configuration denotes by dashed line), *OB* - the working side gnomon with the length *b*, located at an angle $\varphi$ to the horizontal, line *BA* – the plummet with the length *h*, *OE* - gnomon base with long *e*, *OA* – the projection of gnomon onto a plane slab with length *a*, *OL* - the distance from the point *O* of gnomon attachment to the border of gnomon shadow at noon to the summer solstice with length $l_{ss}$.

The projected length *a* of gnomon on the surface of the slab is calculated by formula 4 (Fig. 8):

$$a = l_{ss}/(1 + tg\varphi \cdot tg(\varphi - \delta_{ss})), \qquad (4)$$

where *a* - the length of the projection of the gnomon, $l_{ss}$ - the distance from point O to the boundary of the shadow of the gnomon L, $\varphi$ - latitude of location, $\delta_{ss} = \varepsilon$ - declination of the sun at the summer solstice.

We calculated the length of the projection of the gnomon $a \approx 11.9$ cm. The number of small holes at this distance from the point *O* on the surface of the slab is located parallel groove - scale. It is represented by the dotted line in Figure 4. Breakpoints in the lines correspond to the wells. We assume that the number of wells marked the equinox, because the shadow of the gnomon plummet at sunrise / sunset on the day of the equinox coincided with the next wells. At a distance of ≈7 cm from the point *O* the wells are reviewed at the slab. They could serve for fixing the gnomon at *E* and determined the length of the gnomon base *e* (Fig. 4). The length of gnomon base did not affect the work of analyzed horizontal sundial, and determined only the design features of the gnomon. However, it should be noted, that the length of the gnomon base *e* must always be less or equal to the length of the gnomon projection *a*.

The height of the gnomon apex and its length can be calculated by formulas 5 and 6:

$$h = a \cdot tg\varphi \tag{5}$$

$$b = a/\cos\varphi, \tag{6}$$

where *h* - height of the gnomon apex, *a* - the length of the gnomon projection, *b* - the length of the gnomon (Fig. 8).

Our calculated according to formulas 5 and 6 the values: $h≈13.4$ cm, $b≈17.9$ cm.

The length of the shadow of the gnomon at noon of equinox and the winter solstice[3] calculated by the formulas 7 - 10:

$$LA_{eq} = h \cdot tg(\varphi - \delta_{eq}) \tag{7}$$

$$LA_{ws} = h \cdot tg(\varphi - \delta_{ws}) \tag{8}$$

$$l_{eq} = a + h \cdot tg(\varphi - \delta_{eq}) \tag{9}$$

$$l_{ws} = a + h \cdot tg(\varphi - \delta_{ws}) \tag{10}$$

where $l_{eq}$ - the length of the gnomon shadow at noon of the equinox, $l_{ws}$ - the length of the gnomon shadow at the of winter solstice, *φ* - latitude of location, *LA* - the length of the shadow of the gnomon, excluding the length of its projection, $\delta_{ws} = -\varepsilon$ - declination of the sun at the winter solstice, $\delta_{eq} = 0$ - declination of the Sun on the day of the equinox.

Our calculated according to the formulas 7 - 10 the values: $l_{eq} = 27$ cm, $l_{ws} = 53.8$ cm.

On calculated distances $l_{eq}$ and $l_{ss}$ from point *O*, on the surface of the plate ranks wells are arranged parallel groove - scale. Figure 9 are indicated by dashed lines, where breaks in the lines correspond to the holes.

---

[3] The length of the shadow at noon on the summer solstice in our case has been measured $l_{ss}≈18$ cm

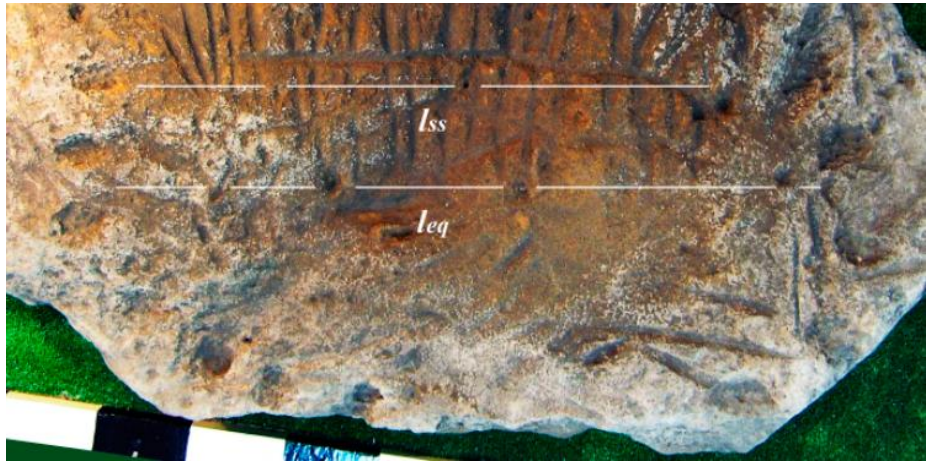

**Figure 9.** Popov Yar-2, tumulus 3, burial 7, plate, side A, the rows of holes: $l_{eq}$ –the equinox, $l_{ss}$ - the summer solstice.

These rows of holes can be used to mark of the equinoxes and summer solstice, like some modern horizontal sundial (Fig. 10). At the Popov Yar-2 plate was used simpler, than modern watches, marking - in the form of straight lines constructed relative to a half-day. The holes on the plate could be independent markers, but could serve to attach, for example, wooden slats, making the marking surface of the plate more visible.

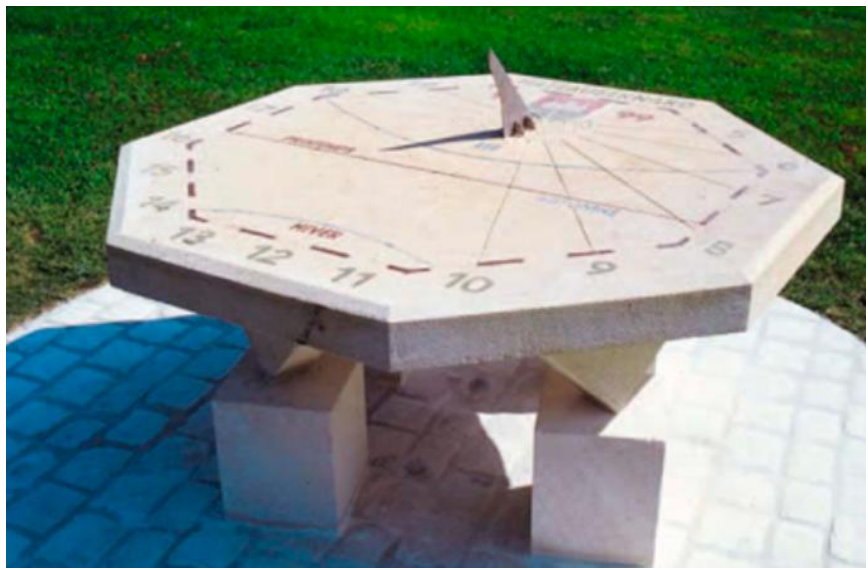

**Figure 10.** Modern horizontal sundial. On the plane with time markings applied hour lines and the lines, label tag the equinoxes and solstices [17].

**Elliptically wells located on the side A (horizontal sundial)**

The wells are located in an ellipse, marked on the side A of Popov Yar-2 slab, except for reduced straight groove with transverse grooves - risks. Line, parallel to reduced straight groove, connecting the opposite wells and passing through the center of the ellipse defined intuitively, was chosen by us as the major axis of the ellipse. The minor axis of the ellipse should be perpendicular to the major axis, and must divide it in half. The point of intersection of the axes is the center of the ellipse. Hour lines of horizontal hours, calculated by the formula 1 (tab. 1, tab. 2) were applied to the corrected photo of side A. Wells in the photo were delineated additionally. Center of the hour lines was placed in the center of the ellipse. As seen in Figure 11, not all of the wells on the side A of Popov Yar-2 slab are consistent with hour lines of horizontal sundial.

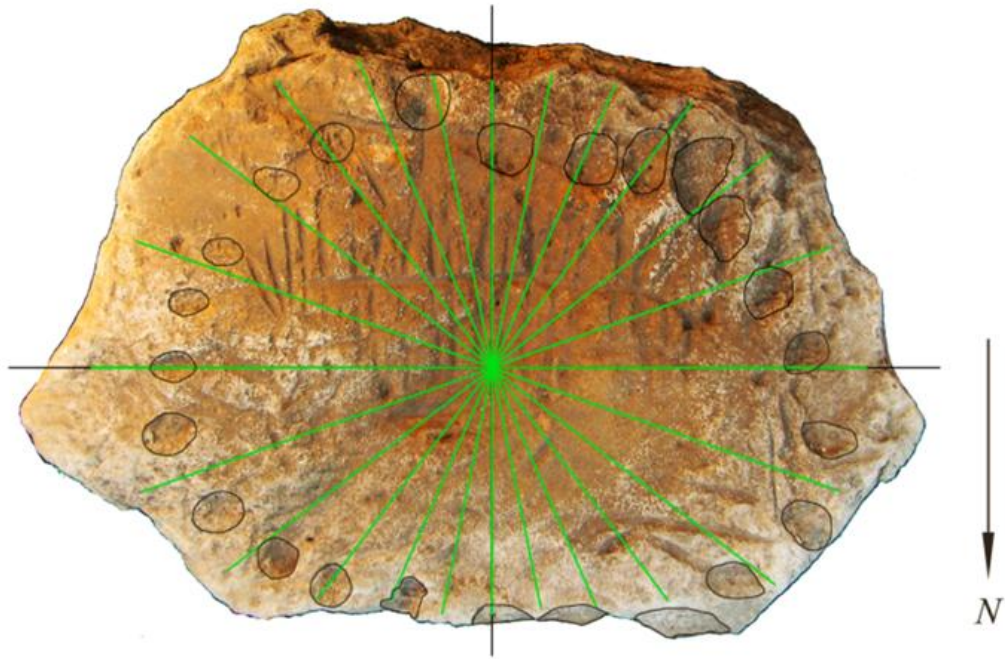

**Figure 11.** Popov Yar-2, tumulus 3, burial 7, slab, side A, hour lines of horizontal sundial (green line).

### Analemmatic sundial of the side A

Elliptical arrangement of the hour marks characteristic for analemmatic sundial, which are believed, to have been invented in the XVII century AD, because they were first described in a while. However, to date known older analemmatic sundial dating from the XVI century AD which are located in Bourg-en-Bresse in France [18]. Thus, the question about time and place of the original appearance of analemmatic sundial is still open. Less perfect and exact prototypes analemmatic sundial whether there were in ancient times are unknown too.

Given that the wells are located in an ellipse on the Popov Yar-2 slab, we hypothesized that they may be the hour marks of analemmatic sundial. The measured semi-major axis of the ellipse (East) $M \approx 32.7$ cm (relative to the center of wells), and the measured semi-minor axis (North), presumably coinciding with the edge of the slab, $m \approx 26.5$ cm (to the inner edge of the well $m \approx 24.5$ cm) for the Popov Yar-2 slab wells. These quantities are related in the analemmatic sundial. Knowing the semi-major axis of the ellipse $M$, using the formula 11, it is possible to calculate the semi-minor axis $m$. For geographic coordinates Popov Yar 2 (latitude 48°26′N and longitude 37°26′E) the calculated semi-minor axis $m$=24,5 cm. It is measured semi-minor axis to the inner edge of the well , which confirms the assumption that the wells on the side A of Popov Yar-2 slab are the hour marks of analemmatic sundial.

Coordinates of the hour marks and the coordinates of the gnomon for analemmatic sundial are calculated as follows [19]:

$$m = M \cdot \sin\varphi \qquad (11)$$

$$x = M \cdot \sin H \qquad (12)$$

$$y = M \cdot \sin\varphi \cdot \cos H \qquad (13)$$

$$Z_{ws} = M \cdot tg\delta_{ws} \cdot \cos\varphi \qquad (14)$$

$$Z_{ss} = M \cdot tg\delta_{ss} \cdot \cos\varphi \qquad (15)$$

$$H' = arctg\left(\frac{tgH}{\sin\varphi}\right) \qquad (16)$$

$$H = 15^0 \cdot (t-12)$$

where $x$ - coordinate of a point on the X axis for analemmatic sundial, $y$ - coordinate of the point on the Y axis for analemmatic sundial, $M \approx 32.7$ cm - measured semi-major axis of the ellipse, $\varphi$ - latitude of location, $H$ - hour angle, $H'$ - angle between the meridian line and the hour line on the sundial, $\delta_{ws} = -\varepsilon$ - declination of the Sun at the winter solstice, $\delta_{ss} = \varepsilon$ - declination of the Sun at the summer solstice, $y = Z_{ws}$ - the winter solstice, $y = Z_{ss}$ - the summer solstice (Fig. 12).

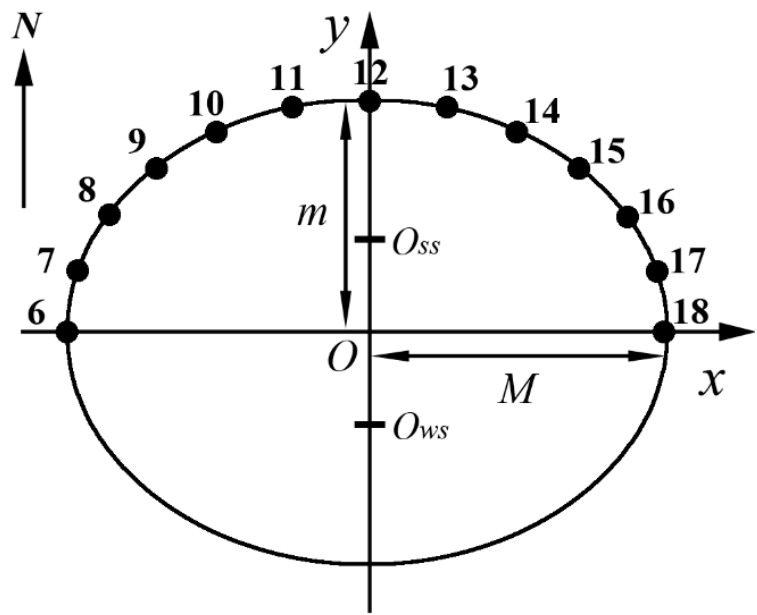

**Figure 12.** The coordinate plane with the hour markers from 6 to 18 hours. $M$ - semi-major axis of the ellipse, m - minor semi axis of the ellipse, $O$ - center of the ellipse, $O_{ws}$ - the position of the gnomon at the winter solstice on analemmatic sundial, $O_{ss}$ - the position of the gnomon at the summer solstice for analemmatic sundial.

The results of our calculations of the coordinates x and y of the hour marks of analemmatic sundial by formulas 12 and 13 are given in Table 3. Calculated according to the formula 14 of gnomon position at the winter solstice $Z_{ws}$=-9.6 cm calculated by the formula 15 of gnomon position at the summer solstice $Z_{ss}$=9.6 cm. The gnomon is at the center of coordinates at the equinox.

Table 3. Coordinates of the hour marks of analemmatic sundial. $H$ – the hour angle of the Sun, $H^/$ - angle between the meridian line and the hour line on the sundial, $t$ - time, $x$ - coordinate of the mark on the axis $X$, $y$ - coordinate of the mark on the axis $Y$.

|  | t, (hour) | | | | | | | | | | | | |
|---|---|---|---|---|---|---|---|---|---|---|---|---|---|
|  | 6 | 7 | 8 | 9 | 10 | 11 | 12 | 13 | 14 | 15 | 16 | 17 | 18 |
| H, (⁰) | -90.0 | -75.0 | -60.0 | -45.0 | -30.0 | -15.0 | 0.0 | 15.0 | 30.0 | 45.0 | 60.0 | 75.0 | 90.0 |
| H/, (⁰) | -90.0 | -78.7 | -66.6 | -53.2 | -37.7 | -19.7 | 0.0 | 19.7 | 37.7 | 53.2 | 66.6 | 78.7 | 90.0 |
| x, (cm) | **-32,7** | -31,6 | -28,3 | -23,1 | -16,4 | -8,5 | 0,0 | 8,5 | 16,4 | 23,1 | 28,3 | 31,6 | **32,7** |
| y, (cm) | 0,0 | 6,3 | 12,2 | 17,3 | 21,2 | 23,6 | **24,5** | 23,6 | 21,2 | 17,3 | 12,2 | 6,3 | 0,0 |

Hour lines of analemmatic sundial, according to the hour angles calculated by the formula 16, were applied to the corrected photo of side A (Fig. 13). Center of the hour lines is the site of the gnomon attachment in the equinox and corresponds to the center of the ellipse - the point *O* with coordinates *(0, 0)*. Analemmatic sundial gnomon is a vertical rod which is moved along the *Y* axis, between the points $O_{ws}$ and $O_{ss}$. Hour lines for equinox highlighted in yellow in Figure 13. Coordinates of the ends of these lines are the coordinates of the calculated hour marks (Table 3). Dotted lines in the range from 18 to 6 hours on Figure 15 are hypothetical hour lines. They can be calculated similarly of the hour lines in range from 6 to 18 hours, and they are arranged symmetrically. Figure 11 shows that the ends of the hour lines coincide with the wells sufficiently good, especially in the operating range from 4 to 20 hours on true solar time[4]. This coincidence also confirms the assumption that the wells on the side A of Popov Yar-2 slab are the hour markers of analemmatic sundial.

For proper operation of analemmatic sundial at the winter solstice the gnomon should be set on the line *WS*, passing through the point $O_{ws}$ *(0, -11.14)* (Fig. 12). The duration of the day in the winter is shorter, so the hour line will cover a smaller range of hours - from 8 to 16 hours, approximately (Fig. 14). In the case of the Popov Yar-2 slab, *WS* line coincides with the scale of horizontal sundial with two gnomons, marked up for the daylight hours of the winter solstice. This coincidence is also indicative of the fact that the side A of Popov Yar-2 slab is analemmatic sundial.

At the summer solstice the gnomon of analemmatic sundial should be set on the line *SS*, passing through the point $O_{ss}$ *(0, 11.14)* (Fig. 12). Duration of the light day in the summer of the biggest, so hour lines include the maximum hourly range - from 4 to 20 hours (Figure 15). Wells appropriate for that time range can be seen at the slab most well, and make most carefully, which also evidence in favor of analemmatic sundial.

---

[4] In the modern life, is mainly used zone time, often does not coincide with the true solar time.

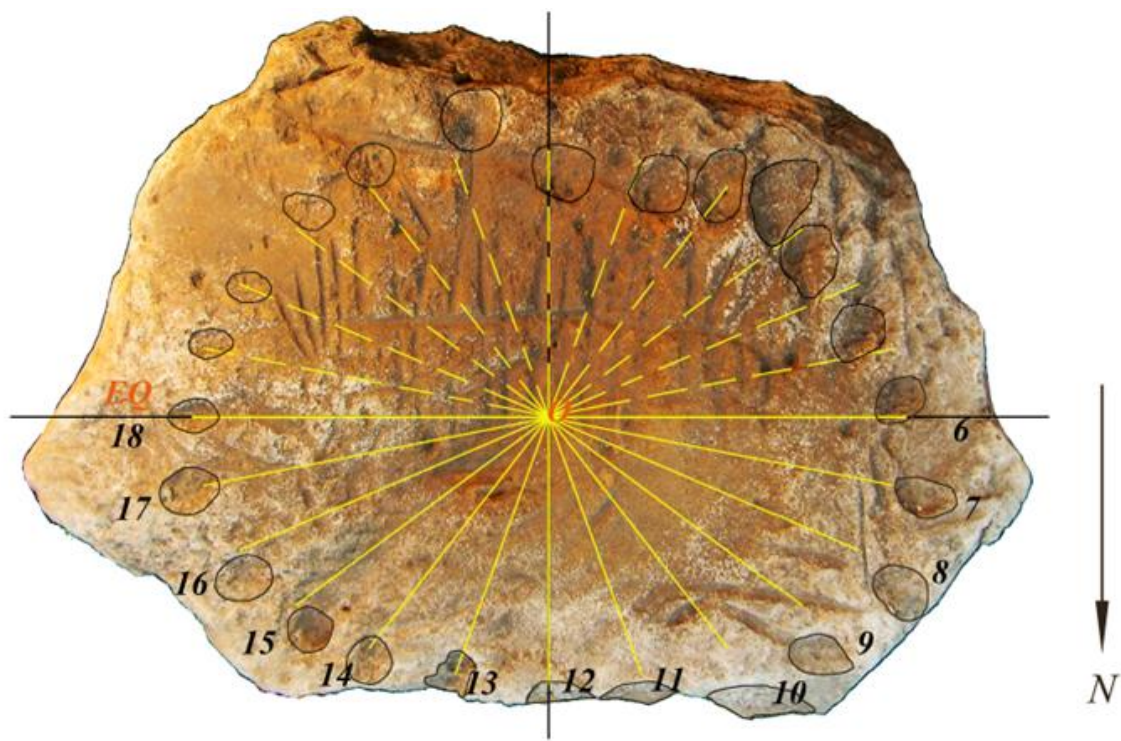

**Figure 13.** Popov Yar-2, tumulus 3, burial 7, slab, side A, hour lines of analemmatic sundial for the equinox (yellow line). Numerals indicate hours according to the hour lines and marks. *EQ* - the line, on which the gnomon should be set at the equinoxes.

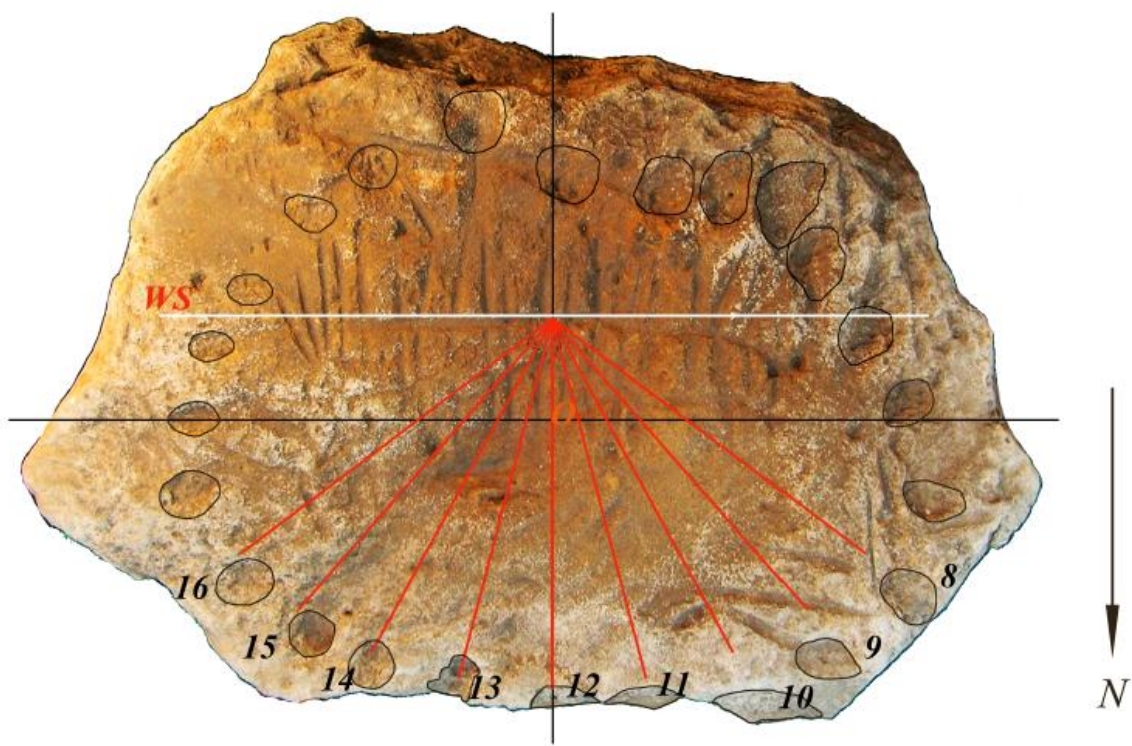

**Figure 14.** Popov Yar-2, tumulus 3, burial 7, slab, side A, hour lines analemmatic sundial for the winter solstice (red lines). Numerals indicate hours according to the hour lines and marks. *WS* – line, on which the gnomon should be set at the winter solstice.

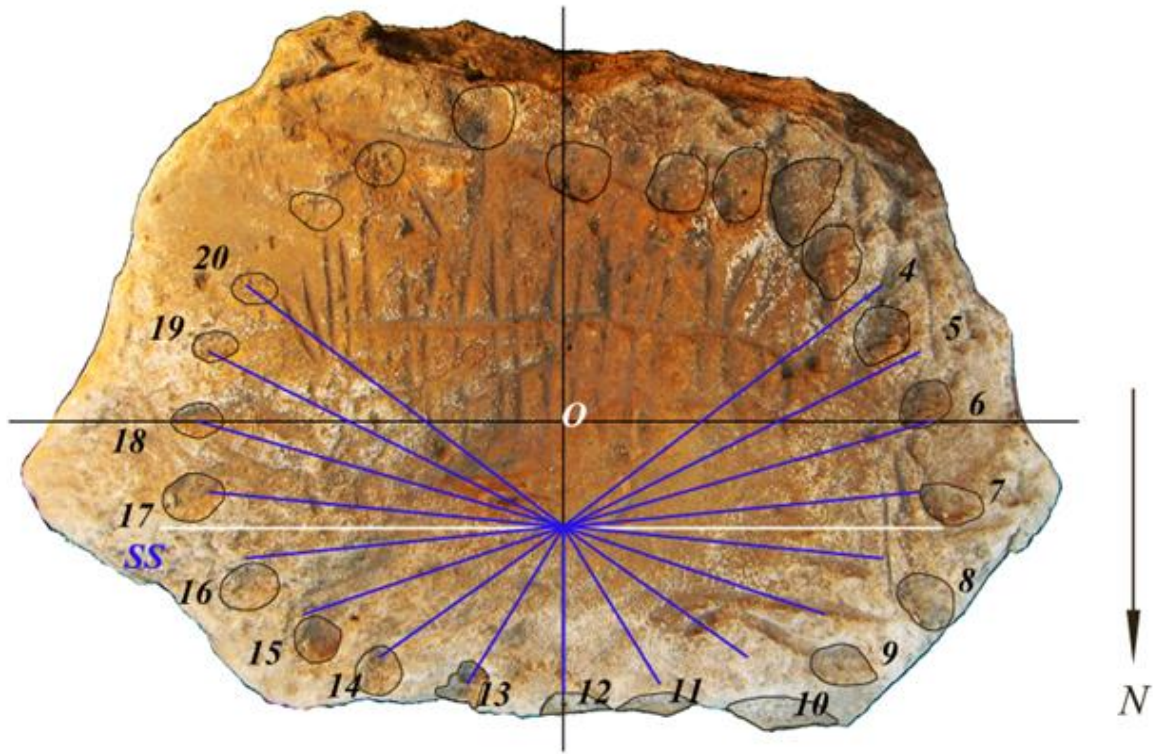

**Figure 15.** Popov Yar-2, tumulus 3, burial 7, slab, side A, hour lines analemmatic sundial for the summer solstice (blue line). Numerals indicate hours according to the hour lines and marks. *SS* - the line, on which the gnomon should be set at the summer solstice.

## Analysis image of the side B

On the surface side B of the Popov Yar-2 slab are viewed in-depth line- groove and the oval from wells, as well as on the side of the A (Fig. 1). In the central part of the oval is the thin groove of the same shape and orientation in space as the groove-scale of horizontal sundial on the side A (Fig. 13).

Long grooves, crossing the central groove-scale on the side of B, most likely, the grooves - risks by analogy with the scale of the horizontal sundial on the side A. The distance between the grooves - risks on the side B twice as much then on the side of A. Therefore, the gnomon on the side B located twice as far from the groove - scale. In this case, places of fastening of gnomons will be out of the slab, which simplifies the process of fixing the gnomons technologically. On the side B grooves - risks mark hours according to the hour lines (Tab. 1, Tab. 2) calculated by the formula 1 (Fig. 16).

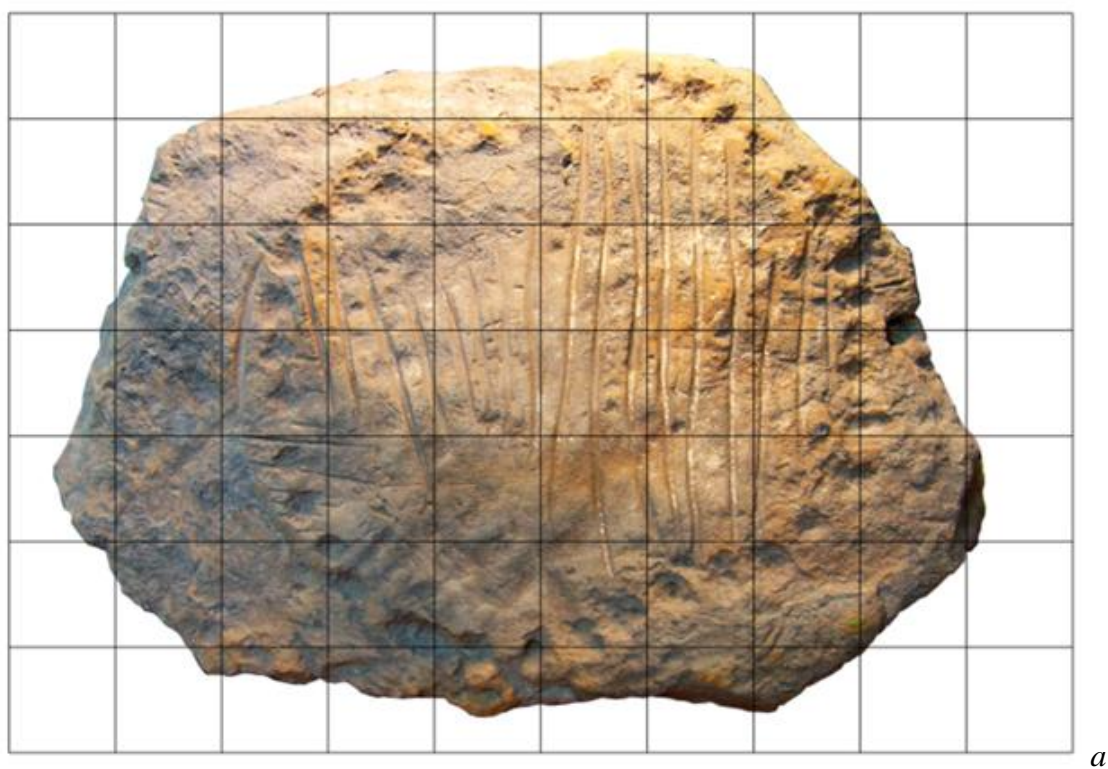

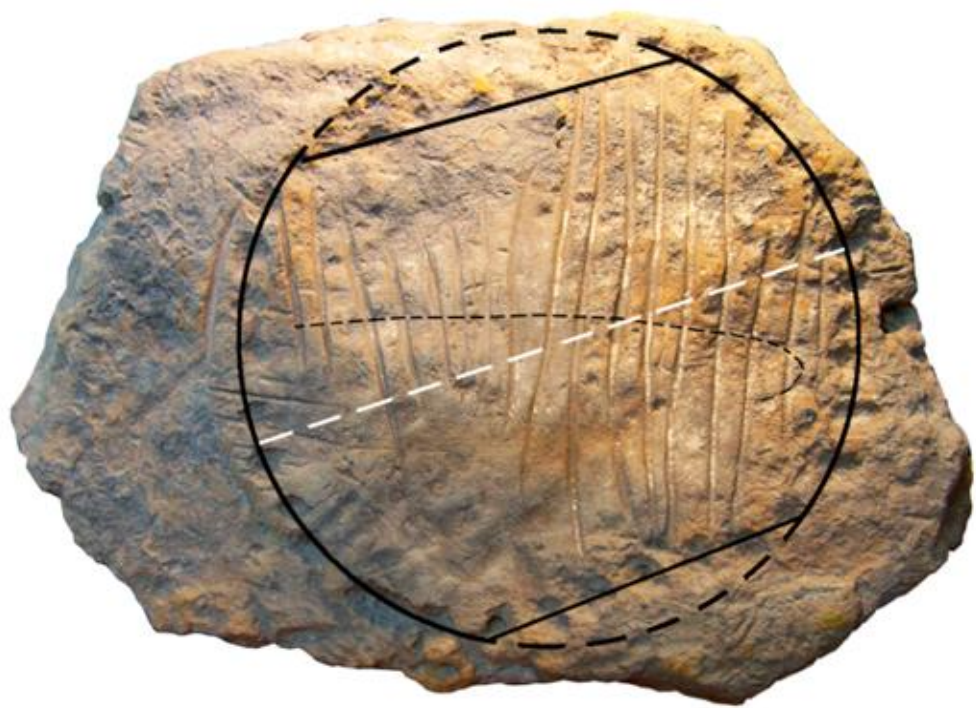

**Figure 16.** Popov Yar-2, tumulus 3, burial 7, slab, adjusted photo of party B: *a* - photo with a large-scale grid, *b* – photo with the drawn circle and secants. In the image is shown in a circle and intersecting approximating the oval from the wells, the symmetry axis of the oval (highlighted in white), the groove-scale of horizontal sundial (marked by the dotted line in the center of the circle).

Hour scale of horizontal sundial on the side B is different from the scale on the side A smaller hour range: 8.5 to 13.5 hours for the gnomon I and from 11 to 16 hours for the gnomon II. At this parameter horizontal sundial on the B-side inferior of analog on side A. Oval formed of wells on the side of B resembles an ellipse, but is not it. It corresponds to the circle with two secants on opposite sides (Fig. 18). The major axis of the oval shape on the side of B passes through the center of the circle, and approximately parallel to the secants. As seen in Figure 18, the major axis of the oval shapes not parallel to the horizontal scale of sundial, as in the case of the major axis on the side A.

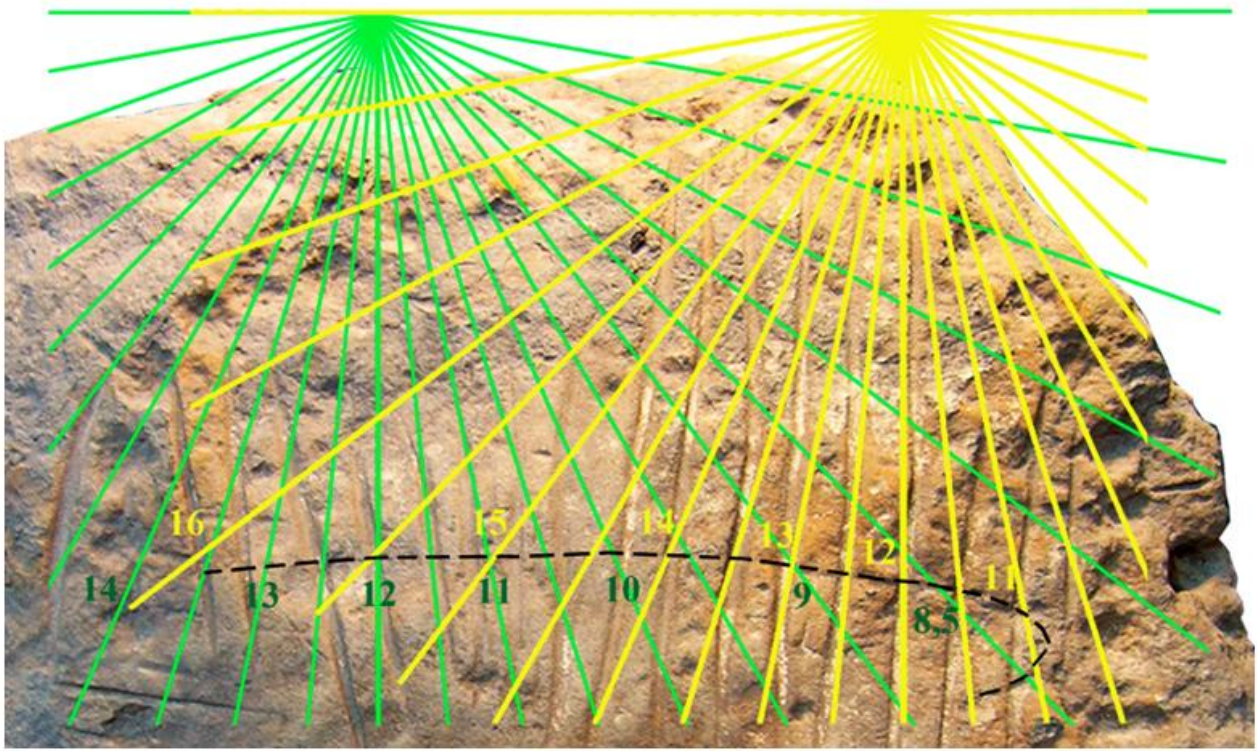

**Figure 17.** Popov Yar-2, tumulus 3, burial 7, slab, side B, hour lines and the scale of horizontal sundial with two gnomons. Numerals indicate hours according to the hour lines.

Almost all of the wells at the side B are close to each other, and many contact their edges. This situation does not correspond to wells horizontal position marks or analemmatic sundial and only creates visibility the markup of sundial. Most likely, the B-side trying to mimic of the layout analemmatic sundial. The need for marking and use of side B could be due to a breakdown of the northern edge of the slab, which led compromise the integrity wells on the side A [20].

The imitation of analemmatic sundial marking on the side B, coupled with less accurate of horizontal sundial than the side A, perhaps indicating a gradual loss accidentally penetrated to the Northern Black Sea coast high-tech knowledge. This loss is likely to have resulted from the lack of stable contacts with the centers of ancient civilizations in the Late Bronze Age at the Srubna culture of the Northern Black Sea coast.

## Analemmatic sundial from the tumulus field Tavria-1

The slab from Srubna burial 2 of tumulus 1 of tumulus field Tavriya-1 (Rostov region, Russia) in the Northern Black Sea coast has wells positioned in an ellipse too (Fig. 18 a).

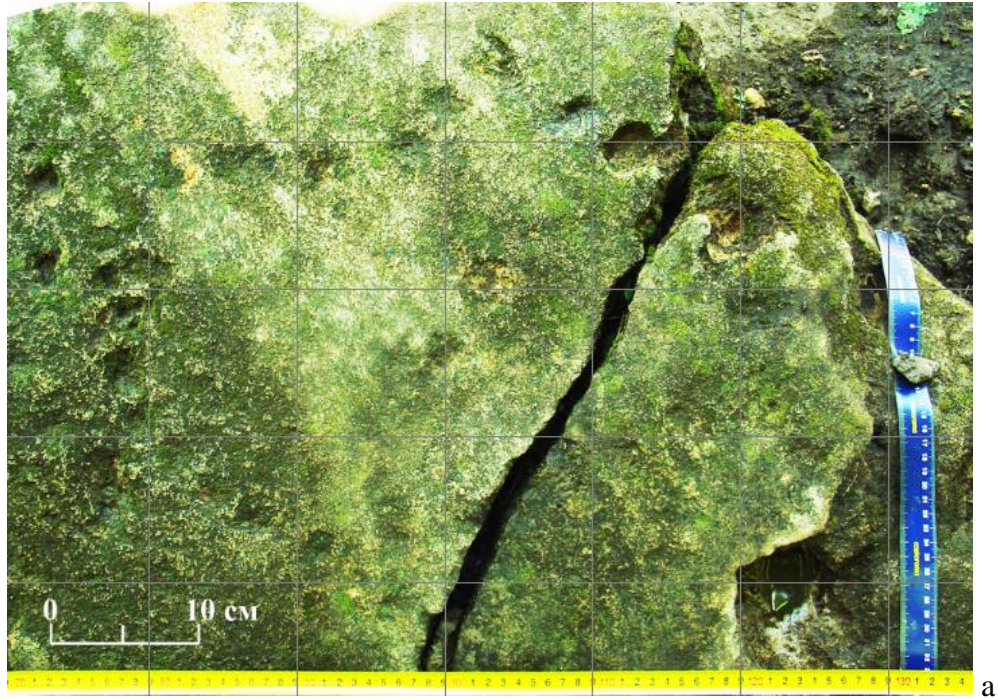

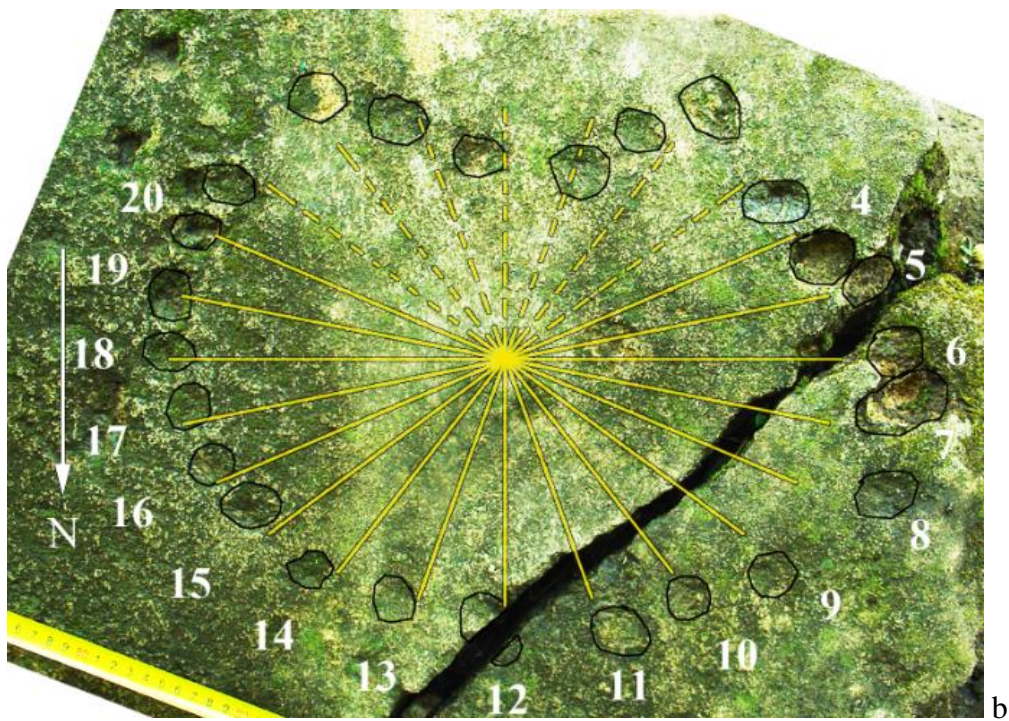

**Figure 18.** Tavriya-1, tumulus 1, burial 2, the slab, a - adjusted photo of slab with wells positioned in an ellipse (fragment); b - hour lines of analemmatic sundial; numerals indicate hours, the appropriate hour lines and marks.

This slab was discovered in 1991 by an archaeological expedition led by P. A. Larenok [21]. At the moment, it is stored in the "Archaeological Museum - Reserve" Tanais "(Rostov region, Russia). The slab is in the open air and it becomes covered lichen, so small details seen badly on its surface. The slab cracked during transport addition to all. Wells were poorly visible in recent years for these reasons. However, we were able to analyze them nevertheless. Calculations for analemmatic sundials were made by the formulas 11-16. The latitude *Lat=47°17' N* for Tavriya-1. The calculation results are presented in Table 4. The measured semi-major axis M ≈ 19 cm (up to a point in the center of the eastern well). The measured semi-minor axis of the ellipse m ≈ 15 cm (up to a point in the center of the northern wells) or m ≈ 13.5 cm (up to the inner edge of the northern well). The calculated m = 14 cm, and equal to measured m, approximately.

Table 4. Coordinates of the hour marks of Tavriya-1 analemmatic sundial. $H$ – the hour angle of the Sun, $H'$ - angle between the meridian line and the hour line on the sundial, $t$ - time, $x$ - coordinate of the mark on the axis $X$, $y$ - coordinate of the mark on the axis $Y$.

|  | t, (hour) | | | | | | | | | | | | |
|---|---|---|---|---|---|---|---|---|---|---|---|---|---|
|  | 6 | 7 | 8 | 9 | 10 | 11 | 12 | 13 | 14 | 15 | 16 | 17 | 18 |
| H, ($^0$) | -90.0 | -75.0 | -60.0 | -45.0 | -30.0 | -15.0 | 0.0 | 15.0 | 30.0 | 45.0 | 60.0 | 75.0 | 90.0 |
| H', ($^0$) | -90.0 | -78.9 | -67.0 | -53.7 | -38.2 | -20.0 | 0.0 | 20.0 | 38.2 | 53.7 | 67.0 | 78.9 | 90.0 |
| x, (cm) | **-19.0** | -18.4 | -16.5 | -13.4 | -9.5 | -4.9 | 0.0 | 4.9 | 9.5 | 13.4 | 16.5 | 18.4 | **19.0** |
| y, (cm) | 0.0 | 3.6 | 7.0 | 9.9 | 12.1 | 13.5 | **14.0** | 13.5 | 12.1 | 9.9 | 7.0 | 3.6 | 0.0 |

Hour lines of analemmatic sundial, in accordance with the hour angles, calculated by the formula 16, were applied to the adjusted photo of the slab (Fig. 18 b).

Seen in the figure 18 b, that the ends of the hour lines coincide with the wells, not bad enough, in the working range of analemmatic sundial from 4 to 20 hours for the true solar time, especially. This coincidence confirms the assumption that the wells on the Tavriya-1 slab are the analemmatic sundial, also. The Tavriya-1 slab has the wells on the surface yet. We plan to devote a separate study for this slab.

Conclusion

Thus, in this study we have proved that the side A of Popov Yar-2 slab is an ancient sundial: horizontal and analemmatic, simultaneously. These are the oldest known sundial in the Northern Black Sea, and analemmatic sundial is the oldest discovered analemmatic sundial in the world.

Analemmatic sundial possible to measure the time in the range from 4 to 20 hours at intervals of one hour. Linear scale of horizontal sundial with two gnomons possible to measure the time from 7.5 to 14.5 hours for the first gnomon, and range from 10.5 to 16.5 hours for the second gnomon. In these ranges scale capable of measuring time every half hour. Reconstruction of the linear parameters of gnomons and scale interval of the horizontal sundial evidence about mediated influence protoscientific knowledge of ancient Egypt to the Srubna population of the Northern Black Sea coast.

In this study, we prove that the Tavriya-1 slab is analemmatic sundial too. Detection of two slabs with analemmatic sundials confirms not random location of wells on the ellipse. This fact indicates the existence of analemmatic sundials during the Bronze Age in the northern Black Sea region is real.

The author expresses deep gratitude to Anatoliy Usachuk and Yuriy Polidovich are the Donetsk archaeologists and employees of Donetsk Museum of Regional Studies, to Vera Larenok and Pavel Larenok are the Rostov archaeologists, to Valeriy Chesnok is employee of "Archaeological Museum – Reserve "Tanais" for the support of this international research.